\documentclass[twocolumn,showpacs,preprintnumbers,amsmath,amssymb]{revtex4}
\input psfig.sty

\usepackage{graphicx}
\usepackage{dcolumn}
\usepackage{bm}

\begin{document}


\title{Constraining the dark energy with galaxy clusters X-ray data}

\author{J. A. S. Lima} \email{limajas@dfte.ufrn.br}

\author{J. V. Cunha} \email{jvital@dfte.ufrn.br}

\affiliation{Departamento de F\'{\i}sica, Universidade Federal do
Rio Grande do Norte, C.P. 1641, 59072-970, Natal, RN, Brasil}

\author{J. S. Alcaniz} \email{alcaniz@astro.washington.edu}

\affiliation{Astronomy Department, University of Washington,
Seattle, Washington, 98195-1580, USA}


\vspace{0.3cm}

\date{\today}

\begin{abstract}
The equation of state characterizing the dark energy component is
constrained by combining Chandra observations of the X-ray
luminosity of galaxy clusters with independent measurements of the
baryonic matter density and the latest measurements of the Hubble
parameter as given by the HST key project. By assuming a spatially
flat scenario driven by a ``quintessence" component with an
equation of state $p_x = \omega \rho_x$ we place the following
limits on the cosmological parameters $\omega$ and
$\Omega_{\rm{m}}$: (i) $-1 \leq \omega \leq -0.55$ and
$\Omega_{\rm m} = 0.32^{+0.027}_{-0.014}$ (1$\sigma$) if the
equation of state of the dark energy is restricted to the interval
$-1 \leq \omega < 0$ (\emph{usual} quintessence) and (ii) $\omega
= -1.29^{+0.686}_{-0.792}$ and $\Omega_{\rm{m}} =
0.31^{+0.037}_{-0.034}$ ($1\sigma$) if $\omega$ violates the null
energy condition and assume values $< -1$ (\emph{extended}
quintessence or ``phantom'' energy). These results are in good
agreement with independent studies based on supernovae
observations, large-scale structure and the anisotropies of the
cosmic background radiation.
\end{abstract}

\maketitle

\section{Introduction}

Recent observations of Type Ia supernovae (SNe Ia) have provided
direct evidence that the universe may be accelerating
\cite{perlmutter}. These results, when combined with measurements
of cosmic microwave background radiation (CMB) anisotropies and
dynamical estimates of the quantity of matter in the Universe,
suggest a spatially flat universe composed of nearly $\sim 1/3$ of
matter (baryonic + dark) and $\sim 2/3$ of an exotic component
endowed with large negative pressure, the so-called
``quintessence". Nowadays, it has been promptly recognized that
the question related to the nature of this dark energy is one of
the most challenging problems of modern astrophysics, cosmology
and particle physics.

The absence of a natural guidance from particle physics theory
about the nature of this dark component gave origin to an intense
debate and many theoretical speculations. In particular, a
cosmological constant ($\Lambda$) -- the most natural candidate --
is the simplest but not the unique possibility; $\Lambda$ is a
time independent and spatially uniform dark component, which is
described by a perfect fluid with $p_v = - \rho_v$. Some other
candidates appearing in the literature are: a decaying vacuum
energy density, or a time varying $\Lambda$-term \cite{ozer}, a
time varying relic scalar field component (SFC) which is slowly
rolling down its potencial \cite{ratra}, the so-called
``X-matter", an extra component simply characterized by an
equation of state $p_{\rm x}=\omega\rho_{\rm x}$ (XCDM)
\cite{turner,tchiba}, the Chaplygin gas whose equation of state is
given by $p= -A/\rho$ where $A$ is a positive constant
\cite{bert}, models based on the framework of brane-induced
gravity \cite{dvali}, among others \cite{mcreation}. For SFC and
XCDM scenarios, the $\omega$ parameter may be a function of the
redshift (see, for example, \cite{efs}) or still, as has been
recently discussed, it may violate the null energy condition and
assume values $< -1$ \cite{leq1}.

In order to improve our understanding of the actual nature of the
dark energy, an important task nowadays in cosmology is to find
new methods or to revive old ones that could directly or
indirectly quantify the amount of dark energy present in the
Universe, as well as determine its equation of state. In this
concern, the possibility of constraining cosmological parameters
from X-ray luminosity of galaxy clusters constitutes an important
and interesting tool. This method was originally proposed by
Sasaki \cite{sasa} and Pen \cite{pen} with basis on measurements
of the mean baryonic mass fraction in clusters as a function of
redshift. A recent application of a new version of this test was
performed by Allen {\it et al.} \cite{allen, allen1} (see also
\cite{ett}) who analyzed the X-ray observations in some relaxed
lensing clusters observed with Chandra spanning the redshift range
$0.1 < z < 0.5$. By inferring the corresponding gas mass fraction
they placed observational limits on the total matter density
parameter, $\Omega_{\rm m}$, as well as on the density parameter,
$\Omega_{\Lambda}$, associated to the vacuum energy density.

In the present paper, by following the methodology presented in
\cite{allen1}, we discuss quantitatively how the observations of
X-ray gas mass fraction of galaxy clusters constrains the cosmic
equation of state describing the dark energy component. In order
to detect the possibility of bias in the parameter determination
due to the imposition $\omega \geq -1$ we have studied two
different cases: the \emph{usual} quintessence ($-1 \leq \omega <
0$) and the \emph{extended} quintessence (also named ``phantom''
energy \cite{leq1}), in which the $\omega$ parameter may assume
values $< -1$. In the former case, a good agreement between theory
and observations is possible if $0.3 \le \Omega_{\rm m} \le 0.35$
($68\%$ c.l.) and $\omega \le -0.55$. These results are in line
with recent analyzes from distant SNe Ia \cite{garn}, SNe + CMB
\cite{efs}, gravitational lensing statistics \cite{class} and the
existence of old high redshift objects (OHRO's) \cite{alcaniz1}.
For \emph{extended} quintessence we obtain $-2.1 \le \omega \le
-0.6$ ($68\%$ c.l.) with the matter density parameter ranging in
the interval $0.27 \le \Omega_{\rm m} \le 0.34$ ($68\%$ c.l.).

This paper is organized as follows. In Section II we present the
basic field equations and the distance formulas relevant for our
analysis. The corresponding constraints on the cosmological
parameters $\omega$ and $\Omega_{\rm{m}}$ are investigated in
Section III. We finish the paper by summarizing the main results
in the conclusion Section.

\section{Basic equations}

For a spatially flat, homogeneous, and isotropic cosmologies
driven by nonrelativistic matter and a separately conserved exotic
fluid with equation of state, $p_{x} = \omega\rho_{x}$, the
Friedman's equation is given by:
\begin{equation}
({\dot{R} \over R})^{2} = H_{o}^{2}\left[\Omega_{\rm{m}}({R_{o}
\over R})^{3} + (1 - \Omega_{\rm{m}})({R_{o} \over R})^{3(1 +
\omega)}\right],
\end{equation}
where an overdot denotes derivative with respect to time and
$H_{o} = 100h {\rm{Km.s^{-1}.Mpc^{-1}}}$ is the present value of
the Hubble parameter.

In order to derive the constraints from X-ray gas mass fraction in
the next Section we shall use the concept of angular diameter
distance, $D_A(z)$. Such a quantity can be easily obtained in the
following way: consider that photons are emitted by a source with
coordinate $r = r_1$ at time $t_1$ and are received at time $t_o$
by an observer located at coordinate $r = 0$. The emitted
radiation will follow null geodesics so that the comoving distance
of the source is defined by ($c = 1$)
\begin{equation}
r_1 = \int_{t_1}^{t_o} {dt \over R(t)} = \int_{R(t)}^{R_o} {dR
\over \dot{R}(t)R(t)}.
\end{equation}
By considering the above equations, it is straightforward to show
that the comoving distance $r_1(z)$ can be written as
\begin{equation}
r_1(z) = \frac{1}{H_oR_{o}}\int_{\frac{1}{1 + z}}^{1} {dx \over x
\sqrt{\Omega_{\rm{m}}x^{-1} + (1 - \Omega_{\rm{m}}) x^{-(1 +
3\omega)}}},
\end{equation}
where the subscript $o$ denotes present day quantities and $x =
{R(t) \over R_o} = (1 + z)^{-1}$ is a convenient integration
variable. The angular diameter distance to a light source at $r =
r_1$ and $t = t_1$ which is observed at $r = 0$ and $t = t_o$ is
defined as the ratio of the source diameter to its angular
diameter, i.e.,
\begin{equation}
D_{A} = {\ell \over \theta} = R(t_1)r_1 = (1 + z)^{-1}R_{o}r(z),
\end{equation}
which provides, when combined with Eq. (3),
\begin{equation}
D_{\rm A}^{\rm{DE}} = \frac{H_o^{-1}}{(1 + z)}\int_{\frac{1}{1 +
z}}^{1} {dx \over x \sqrt{\Omega_{\rm{m}}x^{-1} + (1 -
\Omega_{\rm{m}}) x^{-(1 + 3\omega)}}}.
\end{equation}
For the standard cold dark matter model (SCDM) we set $\omega=0$
in Eq. (3) and the angular diameter distance reduces to
\begin{eqnarray}
D_{\rm A}^{\rm{SCDM}} = \frac{2H_o^{-1}}{(1 + z)^{3/2}}\left[(1 +
z)^{1/2} - 1\right].
\end{eqnarray}

\begin{figure}
\centerline{\psfig{figure=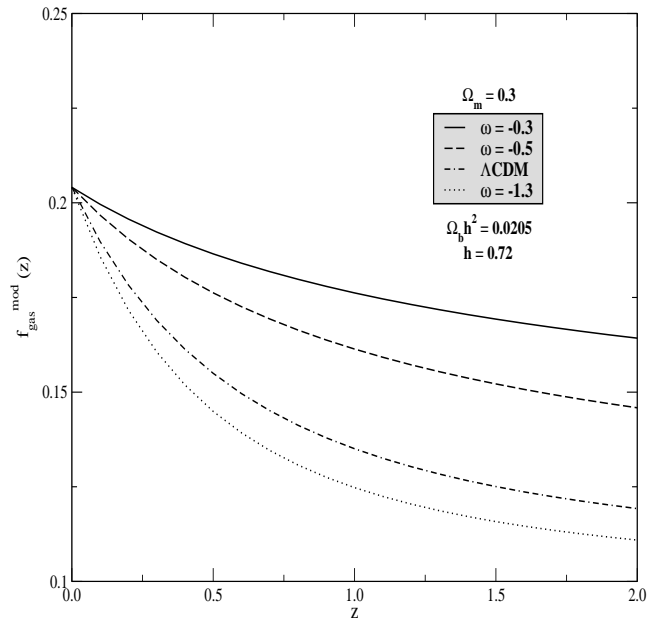,width=3.5truein,height=3.5truein,angle=-90}
\hskip 0.1in} \caption{The model function $f_{\rm gas}^{\rm mod}$
(Eq. 8) as a function of the redshift for selected values of
$\omega$ and fixed values of $\Omega_{\rm{m}} = 0.3$, $\Omega_{\rm
b}h^{2} = 0.0205$ and $h = 0.72$.}
\end{figure}

\section{Constraints from X-ray gas mass fraction}

In our analysis we consider the Chandra data analysed in recent papers by Allen {\it et
al.}\cite{allen, allen1} and Schmidt {\it et al.}\cite{schim}. The
specific data set consists of six clusters distributed over a wide
range of redshifts ($0.1 < z < 0.5$). The clusters studied are all
regular, relatively relaxed systems for which independent
confirmation of the matter density parameter results is available
from gravitational lensing studies. As discussed in Ref.
\cite{allen1}, the systematic uncertainties are $\lesssim 10\%$
(i.e. typically smaller than the statistical uncertainties). The
X-ray gas mass fraction ($f_{\rm gas}$) values were determined for
a canonical radius $r_{2500}$, which is defined as the radius
within which the mean mass density is 2500 times the critical
density of the Universe at the redshift of the cluster. Two data
sets were generated from these data. One in which the SCDM model
with $H_o = 50 {\rm{Km.s^{-1}.Mpc^{-1}}}$ is used as the default
cosmology and the other one in which the default cosmology is the
$\Lambda$CDM scenario with $H_o = 70 {\rm{Km.s^{-1}.Mpc^{-1}}}$,
$\Omega_{\rm{m}} = 0.3$ and $\Omega_{\Lambda} = 0.7$. In what
follows we constrain the basic cosmological parameters using the
SCDM scenario as the default cosmology.

\begin{figure}
\centerline{\psfig{figure=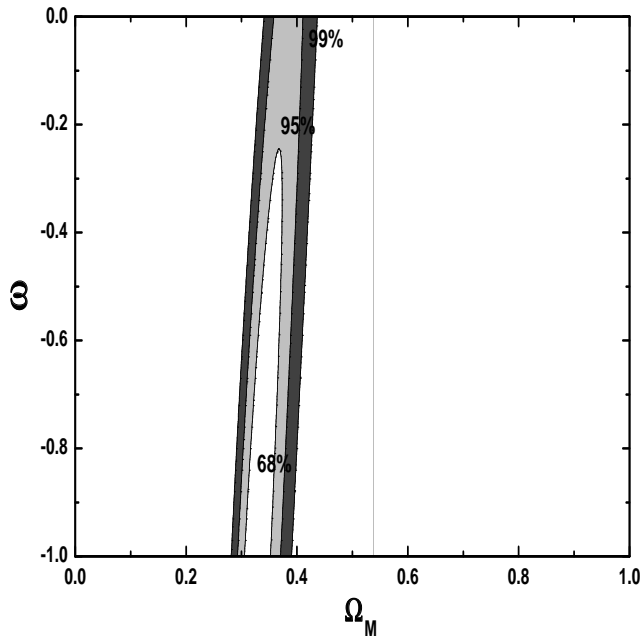,width=3.5truein,height=3.5truein}
\hskip 0.1in} \caption{Confidence regions in the $\Omega_{\rm{m}}
- \omega$ plane by assuming the SCDM model as the default
cosmology. The regions in the graph correspond to 68\%, 95\% and
99\% likelihood contours for flat quintessence scenarios.}
\end{figure}

By assuming that the baryonic mass fraction in galaxy clusters
provides a fair sample of the distribution of baryons at large
scale, the matter content of the universe can be expressed as
\cite{white,hogan}:
\begin{equation}
\Omega_{\rm m} = \frac{\Omega_{\rm b}}{f_{\rm
gas}(1+0.19h^{3/2})},
\end{equation}
where $\Omega_{\rm b}$ stands for the baryonic mass density
parameter. Since $f_{\rm gas} \propto D_{\rm{A}}^{3/2}$
\cite{sasa}, the model function is defined by 
\begin{eqnarray}
f_{\rm gas}^{\rm mod}(z_{\rm i}) = \frac{ \Omega_{\rm b}}
{\left(1+0.19{h}^{3/2}\right) \Omega_{\rm m}} \left[ 2h \,
\frac{D_{\rm A}^{\rm{SCDM}}(z_{\rm i})}{D_{\rm A}^{\rm{DE}}(z_{\rm
i})} \right]^{1.5},
\end{eqnarray}
where the term $(2h)^{3/2}$ represents the change in the Hubble
parameter between the default cosmology and quintessence scenarios
while the ratio ${D_{\rm A}^{\rm{SCDM}}(z_{\rm i})}/{D_{\rm
A}^{\rm{DE}}(z_{\rm i})}$ accounts for deviations in the geometry
of the universe from the SCDM model. Figure 1 shows the behavior
of $f_{\rm gas}^{\rm mod}$ as a function of the redshift for some
selected values of $\omega$ having the values of $\Omega_{\rm b}$
and $h$ fixed. The value of $\Omega_{\rm m}$ is fixed at $0.3$ as
suggested by dynamical estimates on scales up to about $2h^{-1}$
Mpc \cite{calb}. For the sake of comparison, the current favored
cosmological model, namely, a flat scenario with $70\%$ of the
critical energy density dominated by a cosmological constant
($\Lambda$CDM) is also shown.

In order to determine the cosmological parameters
$\Omega_{\rm{m}}$ and $\omega$ we use a $\chi^{2}$ minimization
for the range of $\Omega_{\rm{m}}$ and $\omega$ spanning the
interval [0,1] in steps of 0.02
\begin{eqnarray}
\chi^2 & = &\sum_{i = 1}^{6} \frac{\left[f_{\rm gas}^{\rm
mod}(z_{\rm i})- f_{\rm gas,\,i} \right]^2}{\sigma_{f_{\rm
gas,\,i}}^2}
\end{eqnarray}
where $\sigma_{f_{\rm gas,\,i}}$ are the symmetric
root-mean-square errors for the SCDM data. The $68.3\%$ and
$95.4\%$ confidence levels are defined by the conventional
two-parameters $\chi^{2}$ levels 2.30 and 6.17, respectively.

\begin{figure}
\centerline{\psfig{figure=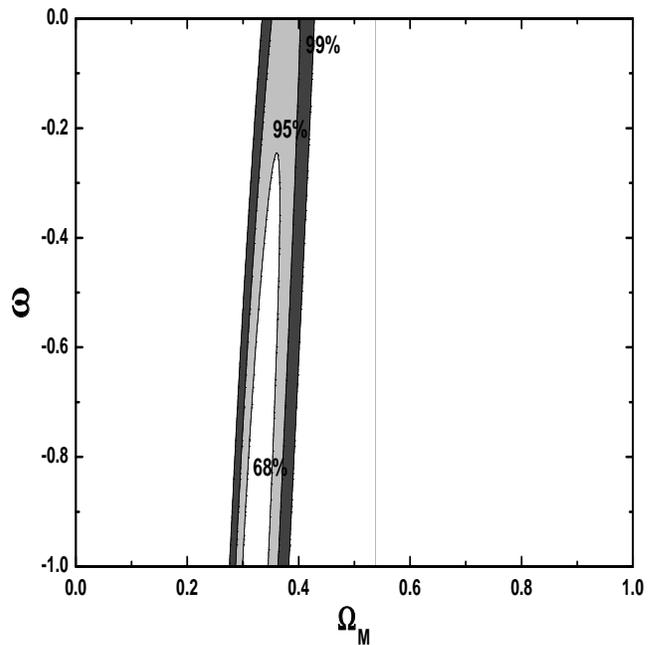,width=3.5truein,height=3.5truein}
\hskip 0.1in} \caption{The same as in Fig. 2 by assuming the
Gaussian priors $h = 0.72\pm0.08$ and $\Omega_{\rm
b}h^2=0.0205\pm0.0018$.}
\end{figure}

In Fig. 2, by fixing the values of $\Omega_{\rm b}$ (0.0205) and $h$ (0.72), we
show contours of constant likelihood (95\% and 68\%) in the
$\Omega_{\rm m}$-$\omega$ plane. Note that the allowed range for
both $\Omega_{\rm m}$ and $\omega$ is reasonably large, showing
the impossibility of placing restrictive limits on these
quintessence scenarios from the considered X-ray gas mass fraction
data. The best-fit model for these data occurs for $\Omega_{\rm m}
= 0.33$ and $\omega = -1.0$ with $\chi^{2} = 1.98$. Such limits
become slightly more restrictive if we assume some {\it a priori}
knowledge of the value of the product $\Omega_{\rm b}h^2 = 0.0205
\pm 0.0018$ \cite{omera} and of the value of the Hubble parameter
$h = 0.72 \pm 0.08$ \cite{freedman}. To illustrate these new
results, in Fig. 3 we show the confidence regions in the
$\Omega_{\rm m}$-$\omega$ plane by assuming such priors. In this
case, the best-fit model occurs for $\Omega_{\rm m} = 0.32$, 
$\omega = -1$ and $\chi^2_{\rm min}=1.95$ with 
the 1-$\sigma$ limits on $\omega$ and $\Omega_{\rm m}$ given,
respectively, by
$$\omega \leq -0.555,
$$ 
and 
$$\Omega_{\rm{m}} = 0.320
^{+0.027}_{-0.014}.
$$ 
For the sake of completeness, we also
verified that by fixing $\omega = -1$ and extending the analysis
for arbitrary geometries the results of \cite{allen1} are fully
recovered.

\begin{figure}
\vspace{.2in}
\centerline{\psfig{figure=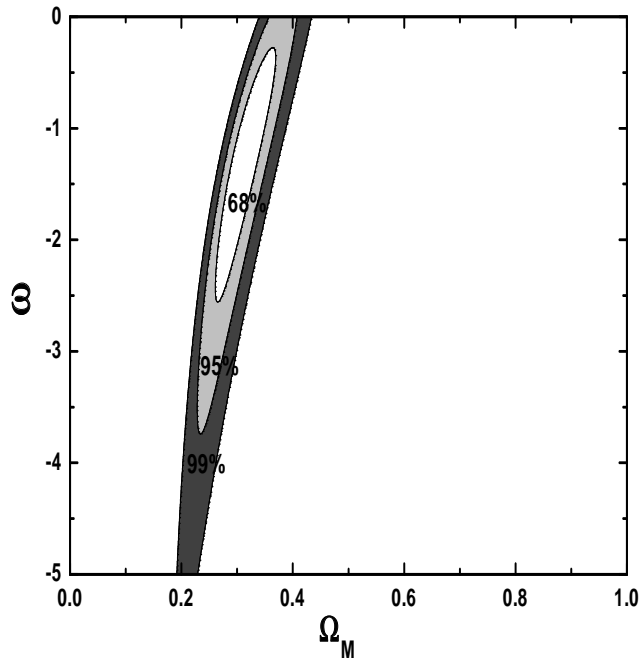,width=3.5truein,height=3.5truein}
\hskip 0.1in} \caption{Constraints on the $\Omega_{\rm{m}} -
\omega$ plane for \emph{extended} quintessence. The regions in the
graph correspond to 68\%, 95\% and 99\% confidence limits. As in
Fig. 3, Gaussian priors on the values of $\Omega_{\rm{b}}h^{2}$
and $h$ were assumed.}
\end{figure}

So far we have assumed that the dark energy equation of state is
constrained to be $\omega \geq -1$. However, as has been observed
recently a dark component with $\omega < -1$ appears to provide a
better fit to SNe Ia observations than do $\Lambda$CDM scenarios
($\omega = -1$) \cite{leq1}. In fact, although having some unusual
properties, this ``phantom" behavior is predicted by several
scenarios as, for example, kinetically driven models \cite{chiba1}
and some versions of brane world cosmologies \cite{sahni} (see
also \cite{mcI} and references therein). In this concern, a
natural question at this point is: how does this extension of the
parameter space to $\omega < -1$ modify the previous results? To
answer this question in Fig. 4 we show the 68\% and 95\%
confidence regions in the ``extended" $\Omega_{\rm{m}}-\omega$
plane by assuming the same {\it a priori} knowledge of the product
$\Omega_{\rm b}h^2$ and of the value of the Hubble parameter as done earlier. From this analysis, we find 
$\Omega_{\rm m} = 0.312 ^{+0.037}_{-0.034}$, $\omega =
-1.29^{+0.686}_{-0.792}$ and $\chi^2_{\rm min}=1.77$ both results
at 1-$\sigma$ level. By assuming no {\it a priori}
knowledge on $\Omega_{\rm b}h^2$ and $h$ we obtain $\omega =
-1.28^{+0.682}_{-0.809}$ while the value of $\Omega_{\rm{m}}$
remains approximately the same. These limits should be compared
with the ones obtained by Hannestad \& M\"ortsell \cite{hann} by
combining CMB + Large Scale Structure (LSS) + SNe Ia data. At
95.4\% c.l. they found $-2.68 < \omega < -0.78$.

\begin{table}
\caption{Limits to $\Omega_{\rm{m}}$ and $\omega$}
\begin{ruledtabular}
\begin{tabular}{lcrl}
Method& Reference& $\Omega_{\rm{m}}$& \quad $\omega$\\ \hline
\hline \\ CMB + SNe Ia:.....& \cite{turner}& $\simeq 0.3$& $\simeq
-0.6$\\
 & \cite{efs}& $\sim$& $< -0.6$\\
SNe Ia....................& \cite{garn}& $\sim$&$< -0.55$\\ SNe Ia
+ GL..........& \cite{waga}& $0.24$& $< -0.7$\\
GL..........................& \cite{class}& $\sim$ & $-0.55$\\ SNe
Ia + LSS.........& \cite{perl}& $\sim$ & $< -0.6$\\
Various...................& \cite{wang}& $0.2 -0.5$& $ < -0.6$\\
OHRO's..................& \cite{alcaniz1}& 0.3&$\leq -0.27$\\
CMB.......................& \cite{balbi}& 0.3& $< -0.5$\\ &
\cite{cora}&\quad $\sim$& $< -0.96$\\ $\Delta
\theta$..........................& \cite{jain}& 0.2-0.4& $\leq
-0.5$\\ $\theta(z)$.........................& \cite{jsa}& 0.2&
$\simeq -1.0$\\ CMB + SNe + LSS.& \cite{bean}& 0.3& $< -0.85$\\
CMB + SNe + LSS.& \cite{hann}& $\sim$& $< -0.71$\\ CMB + SNe +
LSS.\footnote{extended quintessence}& \cite{hann}& $\sim$& $>
-2.68$\\ 
SNe Ia$^{a}$...................& \cite{hann}& 0.45&$ -1.9$\\
SNe + X-ray Clusters $^{a}$..& \cite{schu}& $\simeq
0.29$& $-0.95$\\ X-ray Clusters.......& This Paper
& $\simeq 0.32$& $\leq -0.5$\\ X-ray Clusters $^{a}$.......& This Paper &
$\simeq 0.31$& $-1.29$\\
\end{tabular}
\end{ruledtabular}
\end{table}

At this point we compare our results with other recent
determinations of $\omega$ derived from independent methods. For
example, for the {\it usual} quintessence (i.e., $\omega \geq
-1$), Garnavich {\it et al.} \cite{garn} used the SNe Ia data from
the High-$z$ Supernova Search Team to find $\omega < -0.55$ ($95\%$
c.l.) for flat models whatever the value of $\Omega_{\rm{m}}$
whereas for arbitrary geometry they obtained $\omega < -0.6$
($95\%$ c.l.). Such results agree with the constraints obtained
from a wide variety of different phenomena, using the
``concordance cosmic" method \cite{wang}. In this case, the
combined maximum likelihood analysis suggests $\omega \leq -0.6$,
which rules out an unknown component like topological defects
(domain walls and string) for which $\omega = -\frac{n}{3}$, being
$n$ the dimension of the defect. Recently, Lima and Alcaniz
\cite{jsa} investigated the angular size - redshift diagram
($\theta(z)$) in quintessence models by using the Gurvits'
published data set \cite{gurv}. Their analysis sugests $-1 \leq
\omega \leq -0.5$ whereas Corasaniti and Copeland \cite{cora}
found, by using SNe Ia data and measurements of the position of
the acoustic peaks in the CMB spectrum, $-1\leq \omega \leq-0.93$
at $2\sigma$. More recently, Jain {\it et al.} \cite{jain} used
image separation distribution function ($\Delta \theta$) of lensed
quasars to obtain $-0.8 \leq \omega \leq -0.4$, for the observed
range of $\Omega_m \sim 0.2 - 0.4$ while Chae {\it et al.}
\cite{class} used gravitational lens (GL) statistics based on the
final Cosmic Lens All Sky Survey (CLASS) data to find $\omega <
-0.55^{+0.18}_{-0.11}$ (68\% c.l.). Bean and Melchiorri
\cite{bean} obtained $\omega < -0.85$ from CMB + SNe Ia + LSS
data, which provides no significant evidence for quintessential
behaviour different from that of a cosmological constant. A
similar conclusion was also obtained by Schuecker {\it et al.}
\cite{schu} from an analysis involving the REFLEX X-ray cluster
and SNe Ia data in which the condition $\omega \geq -1$ was
relaxed. A more extensive list of recent determinations of the
quintessence parameter $\omega$ is presented in Table I.

\section{Conclusion}

The determination of cosmological parameters is a central goal of
modern cosmology. We live in a special moment where the emergence
of a new ``standard cosmology'' driven by some form of dark energy seems to be
inevitable. The uncomfortable situation for some comes from the
fact that the emerging model is somewhat more complicated
physically speaking, while for others it is exciting because
although preserving some aspects of the basic scenario a new
invisible actor which has not been predicted by particle physics
is coming into play.

Using the reasonable {\emph{ansatz}} of constant gas mass fraction
at large scale, we placed new limits on the $\Omega_{\rm m}$ and
$\omega$ parameters for a flat dark energy model. The galaxy
cluster data used corresponds to regular, relaxed systems whose
$f_{\rm gas}(r)$ profiles are essentially flat around $r_{2500}$,
the mass results were confirmed from gravitational lensing studies
and the residual systematic uncertainties in the $f_{\rm gas}$
values are small \cite{allen1}. Naturally, the analysis presented
here also reinforces the interest in searching for X-ray data both
for less relaxed clusters, and perhaps more important, at higher
redshifts. Hopefully, our constraints will be more stringent when
further Chandra, XMM-Newton and more accurate gravitational
lensing data for clusters become available near future. In this
concern, we recall that X-ray data from galaxy clusters at high
redshifts and the corresponding constraints for $\Omega_m$ will
play a key hole in the coming years because their relative
abundance (and consequently the value of $\Omega_m$ itself) may
also independently be checked trough the Sunyaev-Zeldovich
effect\cite{C2003}.

As we have seen, the X-ray data at present also favor eternal
expansion as the fate of the Universe in accordance with SNe Ia
data\cite{perlmutter}. Our estimates of $\Omega_{\rm m}$ and
$\omega$ are compatible with the results obtained from many
independent methods (see Table I). We emphasize that a combination
of these X-ray data with different methods is very welcome not
only because of the gain in precision but also because most of
cosmological tests are endowed with a high degree of degeneracy
and may constrain rather well only specific combinations of
cosmological parameters but not each parameter individually. The
basic results combining different methods will appear in a
forthcoming communication \cite{next}.

\begin{acknowledgments}
This work was partially supported by the Conselho Nacional de
Desenvolvimento Cient\'{\i}fico e Tecnol\'{o}gico (CNPq),
Pronex/FINEP (No. 41.96.0908.00), FAPESP (00/06695-0) and CNPq
(62.0053/01-1-PADCT III/Milenio).

\end{acknowledgments}


\end{document}